%% file: Manuscript_draft.tex
\documentclass[10pt,letterpaper]{article}
\usepackage[top=0.85in,left=2.75in,footskip=0.75in,marginparwidth=2in]{geometry}
\usepackage[utf8]{inputenc}
\usepackage[T1]{fontenc}
\usepackage[utf8]{inputenc}
\usepackage{notes2bib}
\usepackage{microtype}
\DisableLigatures[f]{encoding = *, family = * }

\raggedright
\usepackage{changes}
\definechangesauthor[name={TK}, color=blue]{TK}
\definechangesauthor[name=Karina, color=magenta]{KK}
\definechangesauthor[name=Grzegorz, color=green]{GB}
\definechangesauthor[name=Jarek, color=purple]{JM}

\usepackage{changepage}

\usepackage[aboveskip=1pt,labelfont=bf,labelsep=period,singlelinecheck=off]{caption}

\makeatletter
\renewcommand{\@biblabel}[1]{\quad#1.}
\makeatother

\usepackage{amssymb}

\usepackage{color}
\usepackage{url}

\usepackage{lastpage}
\usepackage{fancyhdr}
\usepackage{graphicx}
\fancyheadoffset[L]{2.25in}
\fancyfootoffset[L]{2.25in}

\usepackage{sidecap}
\usepackage{nameref,hyperref}
\usepackage{wrapfig}
\usepackage[pscoord]{eso-pic}
\usepackage[fulladjust]{marginnote}
\usepackage[comma,numbers,super,sort&compress]{natbib}
\reversemarginpar
\usepackage{my_commands}
\begin{document}
\vspace*{0.35in}
\begin{flushleft}
{\Large
\textbf\newline{\input{title}}
}
\newline
\\
\renewcommand{\thefootnote}{\fnsymbol{footnote}}
Tomasz Kalwarczyk\textsuperscript{1,*},
 Grzegorz Bubak\textsuperscript{1},
Jarosław Michalski\textsuperscript{1},
Antoni Lis\textsuperscript{1},
Karina Kwapiszewska\textsuperscript{1},
Marta Pilz\textsuperscript{1},
Adam Mamot\textsuperscript{2\footnotemark[2]},
Olga Perzanowska\textsuperscript{2,\footnotemark[3]},
Joanna Kowalska\textsuperscript{2},
Jacek Jemielity\textsuperscript{2},
Robert Hołyst\textsuperscript{1},
\\

\bigskip
\bf{1} Institute of Physical Chemistry, Polish Academy of Sciences\\
Kasprzaka 44/52 01-224, Warsaw, Poland.
\\
\bf{2} Centre of New Technologies, University of Warsaw, Banacha 2c Street, 02-097, Warsaw, Poland
\\
\bigskip
* tkalwarczyk@ichf.edu.pl
\footnotetext[2]{Current address: Max Planck Institute of Biochemistry, Department of Cellular and Molecular Biophysics, Am
Klopferspitz 18, 82152 Martinsried, Germany}
\footnotetext[3]{Interdisciplinary Nanoscience Center, Aarhus University, Gustav Wieds Vej 14,  8000 Aarhus C, Denmark}

\end{flushleft}

\section*{Abstract}
\input{abstract}

\renewcommand{\thefootnote}{\fnsymbol{footnote}}
\section*{Introduction}
\input{intro}
\section*{Materials and methods}
\input{Methods}

%
\input{methodology}
\input{Software_description}
\section*{Results}
\subsection*{Validation and tests}
\input{validation}
\subsection*{Mapping the mRNA concentration in living cell}
\input{Experimental_example}

\section*{Conclusions}
\input{Conclusions}

\section*{Acknowledgments}
Research funded by the Polish Science Fund within the framework of the Virtual Research Institute; grant
WIB-1/2020-O11 - WIB\_HERO. Authors declare no conflict of interests.

\section*{Data availability}
The data that support the findings of this study are deposited
in RepOD repository (\url{https://repod.icm.edu.pl/}) and will be made publicly available upon publication
of the final article.

\section*{Code availability}
The smICA code presented in this article is available at \url{https://doi.org/10.5281/zenodo.20134967} (version
v2.1.1)\cite{smICA_v2_1_1}
The development repository is hosted on GitHub (\url{https://tkmist.github.io/smICA/}.)
\section*{Authors contribution}

\textbf{TK}: Methodology, Software development, Validation, Formal analysis, Investigation, Data
Curation, Visualisation, Supervision, Writing - Original Draft, Writing - Review \& Editing.
\textbf{GB}: Investigation, Data Curation, Methodology, Validation, Writing - Original Draft, Writing - Review \& Editing.
\textbf{JM}: Investigation, Visualisation, Data Curation, Writing - Review \& Editing.
\textbf{AL}: Methodology, Software development, Writing - Review \& Editing.
\textbf{KK}: Investigation, Data Curation, Methodology, Software (testing), Writing - Original Draft, Writing - Review \& Editing.
\textbf{MP}: Methodology, Software (testing), Writing - Review \& Editing.
\textbf{AM}: Resources.
\textbf{JJ}: Resources, Funding acquisition.
\textbf{RH}: Funding acquisition, Writing - Review \& Editing.
%


\bibliography{PTU_CONC_references}

\bibliographystyle{unsrtnat}
\end{document}

%% file: title.tex
smICA: Open-Source Software for Quantitative, Lifetime-Resolved Mapping of Absolute Fluorophore Concentrations in Living Cells

%% file: abstract.tex
Advanced microscopy techniques are essential in biomedical research for visualising and tracking biomolecules within
living cells and their compartments. Conventional fluorescence microscopy methods, however, often struggle with
accurately measuring the absolute concentrations of fluorescent probes in living cells. To overcome these limitations, we
introduce an open-source analysis tool, smICA (Single-Molecule Image to Concentration Analyser). The smICA method offers
quantitative mapping of absolute fluorophore concentrations, lifetime-resolved filtering methods of the signal,
intensity-based cell segmentation, and requires only a few photons per pixel. Our approach also reduces the time required
to determine the mean concentration per cell compared to the standard FCS measurement performed in multiple posts. To
highlight the robustness of the method, we validated it against standard fluorescence correlation spectroscopy (FCS)
measurements by performing \textit{in vitro} (polymers in aqueous solution) and \textit{in vivo} (polymers and EGFP in
living cells) experiments. Finally, we present exemplary studies on the time evolution of fluorescently labelled mRNA
concentration in living cell. The presented methodology, along with the software, is a promising tool for quantitative
single-cell studies, including, but not limited to, protein expression, biomolecule degradation (such as proteins and
mRNA), and monitoring enzymatic reactions.

%% file: intro.tex
\indent The development of advanced microscopy techniques is essential for a quantitative understanding of the processes occurring
in living cells. Microscopy is crucial for the visualisation of cells, the localisation and tracking of single biomolecules, or
the analysis of their concentration. Fluorescence microscopy, particularly confocal microscopy, has become a standard tool in
biomedical research. Numerous techniques have evolved from standard laser scanning confocal microscopy (LSCM), including, but not
limited to, fluorescence correlation
spectroscopy (FCS),\cite{Magde1972, Elson1974, Ehrenberg1974, Koppel1974}  raster image correlation spectroscopy
(RICS),\cite{Brown2008} and fluorescence lifetime imaging microscopy (FLIM). These techniques are based on time-correlated single
photon counting (TCSPC), which enables the tracking of photons emitted by fluorescent molecules, one by one, in time.  \\
\indent There are various methods for monitoring the concentration of fluorescent molecules. Qualitatively, under typical imaging
conditions with a low fluorophore concentration and assuming that the photophysical and environmental parameters remain constant,
the concentration in the given region of the sample can be estimated from the fluorescence intensity, which is directly
proportional to concentration.
Obtaining quantitative results, however, is more challenging, and FCS is the first-choice technique. The disadvantage of the
correlation-based methods is that they are limited to mobile molecules and are blind to molecules that are, for some reason,
immobile. Moreover, the classic FCS measurements provide information from only a single spot. The representative FCS measurements
of concentration can only be conducted under the assumption of homogeneous analyte distribution within the sample, or by probing
several spots within the cell, which is a time-consuming process. Although there have been attempts to perform multipoint
FCS,\cite{Ohsugi2009, Galland2011, Yamamoto2018} which to some extent takes into account the spatial heterogeneity of the samples,
the technique requires customised equipment and is not easily implemented on commercially available setups. The limitations of the
FCS technique and its derivatives, including the restriction to mobile species, spatial constraints, and compartmentalisation,
make absolute concentration quantification in living cells particularly challenging. This difficulty arises because target
biomolecules often localise within specific subcellular
structures and organelles, resulting in an uneven distribution throughout the cell. Mapping the concentrations or analysis of the
distribution of the probe's concentration is more desirable in such cases.\\

\indent While new methods and tools for quantitative microscopy are constantly developed,\cite{Aguilera2025} there are
only a few attempts in the literature to implement quantitative imaging of concentrations in optical
microscopy.\cite{Sugiyama2005, Vukojevic2008, Digman2008} An interesting example of a quantitative approach to studying the number
of fluorescent molecules registered on a given image was proposed by Vukojevic et al.\cite{Vukojevic2008} The authors used
Poissonian statistics
to measure the average number of molecules in the observation volume without temporal or spatial correlation of the images. Due to
strong deviations from the Possonian statistics lying on its ground, the method is limited to two orders of magnitude of the
analyte concentration (1-100 nM). The method described by Vukojevic et al.\cite{Vukojevic2008} has the same theoretical basis as
the Number and Brightness (N\&B) method proposed by the Gratton group,\cite{Digman2008} which utilises the relationship between
the mean and variance of the fluorescence signal to determine the molecular brightness and number of fluorescent molecules. The
advantage of the N\&B is that the molecular brightness of the molecule of interest does not need to be known a priori, as it is
determined from the fluorescence fluctuation signal. Moreover, the method discriminates between the mobile and immobile molecule
fractions.
Another quantitative approach relies on determining the calibration curve of concentration versus intensity, or vice
versa.\cite{Brinkenfeldt2023, Lo2015} Among others, an elegant approach was introduced by Politi et al.,\cite{Politi2018} where
the concentration of fluorescent protein was determined based on the calibration curve of concentration vs. intensity. Next, the
number of proteins was calculated based on concentration, and the effective focal volume was determined from FCS. To determine the
absolute concentrations of reduced nicotinamide adenine dinucleotide (NADH) in cells,\cite{Ranjit2018} Gratton's group proposed
the phasor analysis\cite{Digman2008a} method that successfully addresses the autofluorescence of biological samples and the
crosstalk from other fluorescent dyes. However, the main disadvantages of this method are the requirement of the analyte
concentration calibration and a large number of photons to be registered to perform the analysis. In the phasor analysis method, a
TCSPC histogram of fluorescence lifetimes is created for each pixel of the image, which requires between a few hundred and
thousands of photons per pixel. Therefore, the technique is limited either to highly concentrated samples with high molecular
brightness of the tracers or to the acquisition time, which is elongated due to integration over a large number of frames.\\
\indent To address the aforementioned challenges in determining fluorophore concentration, we previously combined a laser-scanning
confocal microscope with time-correlated single-photon counting (TCSPC) and applied it to quantify transferrin
uptake.\cite{Pilz2025} Here, we extend this approach and present a practical, validated framework for determining the absolute
concentrations of fluorescent tracers, a comprehensive workflow, all combined in a dedicated software package (smICA,
Single-Molecule Image to Concentration Analyser). We also implemented systematic method validation in the workflow
outlined.\cite{smICA_v2_1_1}.\\
\indent Rather than introducing a fundamentally new measurement principle, the proposed approach integrates FLIM/confocal imaging
and TCPSC data acquisition, supported by the FCS-based calibration. As a result, we propose a robust workflow that enables
spatially resolved mapping of absolute concentration under conditions where standard methods such as fluorescence correlation
spectroscopy (FCS) are limited.\\
\indent The presented method does not require a calibration curve for the analyte. Instead, it relies on calibrating the focal
volume using single-point FCS measurements with a standard fluorophore and determining the molecular brightness of the analyte in
the sample of interest using standard FCS measurements. The approach should be interpreted as an extension of FCS-based
quantification into the imaging domain.\\
\indent The smICA software provides: \textit{i)} lifetime-resolved filtering methods to remove unwanted photons originating from
samples’ autofluorescence or dye-dye crosstalk, \textit{ii)} intensity-based cell segmentation to identify the region of interest,
and \textit{iii)} operation in very low photon-count regimes, the statistical reliability of which is analysed in detail as a
function of experimental parameters. Finally, the graphical user interface is constructed on the Dear PyGUI engine, making smICA
simple, intuitive, fast, and cross-platform.\\
\indent The primary focus of this paper is the validation of the proposed framework and dedicated software against the
standard FCS method under varying experimental conditions and with changing photon filtration procedures. We demonstrate
smICA's robustness by utilising measurements in aqueous solutions of fluorescently labelled polymers and in living cells.
Finally, to strengthen the applicability of the smICA workflow, we provide exemplary measurements of fluorescently
labelled mRNA degradation in living cells.

%% file: Methods.tex
\subsection*{Cell culture}
Human lung carcinoma cells (A549 line) were obtained from the American Type Culture Collection (ATCC). The
cells were cultured in DMEM (Institute of Immunology and Experimental Technology, Wrocław, Poland)
supplemented with 10\% fetal bovine serum, L-glutamine (2 mM), penicillin (100 $\mu$g/ml), and streptomycin
(100 $\mu$g/ml) from Sigma-Aldrich. The culture was maintained at 37°C in a 5\% $CO_2$ humidified atmosphere.
Passaging was carried out at approximately 80\% confluence every 2-3 days using 0.25\% Trypsin-EDTA solution
(Sigma-Aldrich) and PBS (Sigma-Aldrich).

\subsection*{mRNA microinjections}

The A549 cells for microinjections were plated 1 or 2 days before the measurements on a 35 mm glass-bottom
cell culture dish (glass coverslip bottom, 0.17 µm, ibidi, Germany) to grow to approximately 50-70 \%
confluence. Microinjections were performed by the Femtojet system (Eppendorf) using disposable sterile
injection glass capillaries (Femtotip II, Eppendorf, 0.5 µm inner and 0.7 µm outer diameter). The injection
time was set to 0.1 s. Before FCS measurements but after microinjections, the growth medium in the culture
dish was washed out and replaced with IMDM medium (Gibco), which is dye-free and characterised with low autofluorescence.

\subsection*{Experimental setups}
The measurements were performed using the NIKON C1 and NIKON A1 confocal microscopes. Both setups are
additionally equipped with the TCSPC acquisition unit (LSM Upgrade-kit from PicoQuant, Germany). The LSM units
are composed of three pulsed diode lasers (485, 561, and 641 nm of wavelength) controlled by the PDL 828
"SEPIA II" controller, the TCSPC module PicoHarp 300, and a pair of single photon avalanche diode (SPAD)
detectors with appropriate filters. We used the NIKON PLAN APO water immersion objective x60 NA 1.20 in all
experiments. Both confocal systems are equipped with an incubator, allowing for temperature control with accuracyy of
0.5\textcelsius.
Measurements in buffers were performed at 25\textcelsius, while experiments with cells were conducted at
36\textcelsius.
\subsection*{Materials}
The fluorescently labelled mRNA (Cy5-m7GRNAegfp, stained with sulfo-Cy5 at the 5' Cap) was synthesised and
purified according to the previously published procedure\cite{mamot2022ethylenediamine}. Briefly, mRNA was in vitro transcribed in
the presence of azido-modified cap analogue (N3-m7GpppG) and subsequently isolated with spin columns. After labelling with
sulfo-Cy5-DBCO, the product was isolated with HPLC and characterised with gel electrophoresis. The non-fluorescent
mRNA
coding for the EGFP was synthesised with the same method using trinucleotide cap-1 (m7GpppAmpG).\cite{mamot2022ethylenediamine}
After synthesis, the mRNA concentration was determined using a NanoDrop. The TRITC-labelled dextran with a molecular weight of 155
kg/mol was used in aqueous solution experiments. In living cells, we used dextran with a molecular weight of 40 kg/mol. Both
polymers were bought from Sigma-Aldrich.

\subsection*{Fluorescence correlation spectroscopy - calibration, brightness and concentration}
Before each experiment, calibration measurements were performed using the fluorescence correlation
spectroscopy technique (FCS). The FCS data were acquired using the SymPhoTime64 software (PicoQuant) and
analysed using the open-source FcsIT software.\cite{kalwarczyk_2026_19367803, FcsIT_Kalwarczyk2026_arXiv} Depending on
the laser and the fluorophore under study, we performed
calibration using aqueous
solutions of
Rhodamine 110 (for  485 nm laser line), or Rhodamine B (561 nm laser line). The diffusion coefficients of dyes at
25\textcelsius\ were taken from the reference \citenum{Kapusta2010}. For
measurements in living cells, the
calibration measurements were performed at 36\textcelsius, and the diffusion coefficients were recalculated,
including the change in temperature and viscosity. From the calibration
measurements, we get the structure parameter $\kappa$ (fixed in target
FCS measurements), the width of the focal volume, $\omega$, and the focal volume, $\Veff$.
The molecular brightness of the probes, $B$, was determined under the same conditions as the final measurement; same
power of the laser and same environment (water for \textit{in vitro} measurements and the living cell cytoplasm).\\
\indent First, we obtained the average number of molecules in the confocal focus, $\Nm$, from analysing the FCS autocorrelation curve and photon-counts per second, $\cntrate$, from the raw FCS time-trace signal. To determine the $\cntrate$ for a given FCS measurement, we calculated
the cumulative number of detected photons for each second of measurement, $\Npt$ and fitted the result with the linear function $\Npt = \cntrate t$. The molecular brightness is given as
$B = \cntrate/\Nm$. For each sample, we performed a series of FCS measurements and calculated an average value of $B$ from the series.\\
\indent FCS measurements in cells were performed in the cell cytoplasm, trying to avoid the endoplasmic reticulum. The
stripped region of the cell marked in the Figure
\ref{fig:fcs_region} depicts the region of the cell where typical FCS measurements were conducted.

\input{figure_1}

%% file: figure_1.tex
\begin{figure}[t!]
 \includegraphics[width=0.8\linewidth
 ]{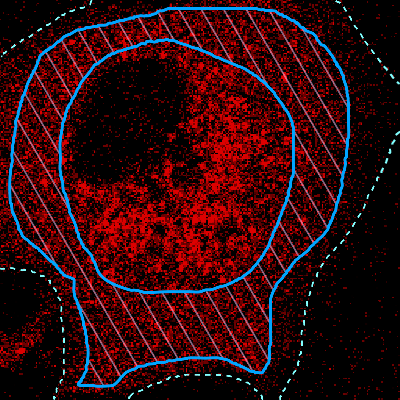}
 \caption{\label{fig:fcs_region} \textbf{Localization of the FCS measurement spots.} The LSCM image of the cell with
fluorescently-labelled (Cy5) mRNA injected into the cytoplasm marked as hashed region marked in the figure corresponds to the
locations where FCS measurements were typically performed. The dashed curve represents the cell boundary. The FCS measurements
were chosen to omit the endoplasmic reticulum and nucleus. }
\end{figure}

%% file: methodology.tex
\subsection*{Quantitative concentration imaging methodology}
\input{figure_2}

\indent The single-photon fluorescence imaging data can be affected by artefacts originating from samples, such as dye crosstalk
and autofluorescence. To conduct quantitative analysis of fluorescent probe concentration, it is
essential to identify each source of
artefact and minimise its impact on the experimental data. Our strategy is described below.\\

\indent The scheme of the methodology is depicted in Figure \ref{fig:met_scheme}.
 First, we use standard FCS measurements to calibrate the confocal volume and determine the molecular
brightness, $B$, of the fluorophore used in further experiments. The molecular brightness depends on the power
of the incident light and the experimental setup settings. Therefore, measuring under the same conditions as
the final experiments is recommended. We further assume that in the sample, $B$ does not change, neither
spatially nor temporally.\\
\indent After calibration, we perform the LSCM imaging combined with the SPAD detectors utilising the TSCPC
method. Standard LSCM hardware provides spectral separation via optical filters and dichroic mirrors. The TCSPC method
used in the time-tagged time-resolved (TTTR) mode,\cite{Wahl2014} additionally allows filtering fluorescence
data by methods
relying on the fluorescence decay pattern, such as fluorescence lifetime filtering (commonly used in
fluorescence lifetime correlation spectroscopy)\cite{enderlein2005using,kapusta2007fluorescence} or the pulsed
interleaved excitation combined with the time gating method. \\
The raw TTTR data are extracted from the binary \emph{.ptu} files, optionally filtered, integrated over all
registered frames, and combined into the image. Each image pixel contains a total number of photons registered
at a given spot in the sample. Next, for each pixel, we calculate the effective number of
detected molecules, $\Nm$, using the fluorophore's molecular brightness, $B$. The mean
number of molecules detected at a given spot is given as $\Nm=\Np\slash\left(B
\tpd\Nf\right)$, where $\Np$ is the mean number of photons detected in the given pixel, $\tpd$ stands for the pixel
dwell time and $\Nf$ is the number of frames acquired.
Finally, knowing the dimensions of the effective focal volume $\Veff$, from the calibration measurements, we calculate
the mean concentration per pixel;
$\Cim = \Nm\slash\left( \Na \Veff \right)$, where
$\Na$ is the Avogadro constant.\\

%% file: figure_2.tex
\begin{figure}[t]
 \includegraphics[width=\linewidth
 ]{Figure_2}
 \caption{\label{fig:met_scheme}\textbf{The workflow scheme for the concentration imaging method.} First, we
calibrate
the FCS setup using a dye with a known diffusion coefficient. From calibration measurements, we get $\omega$ - the width of the
focal volume. Next, we performed the FCS measurements on the target samples to find the molecular brightness, $B$. Finally, the raster
imaging was performed using the laser scanning confocal microscopy equipped with the SPAD detectors. The raw data, in the form of
a time trace of photons acquired during scanning, were further filtered to remove unwanted
background, afterpulsing, or autofluorescence photons. We used fluorescence lifetime-resolved filters that are typicaly used
for the fluorescence lifteime corrlation spectroscopy (FLCS) described in references
\citenum{enderlein2005using} and \citenum{kapusta2007fluorescence}. The filtered signal was combined into frames and integrated
over all
frames. Knowing the $\omega$ and $B$, we calculated the mean concentration of
fluorophores in each pixel averaged over the data acquisition time.}
\end{figure}

%% file: Software_description.tex
\subsection*{The graphical user interface}
The tool for concentration imaging is designed to be as intuitive as possible. We aim to use it by users who are not
trained in
programming
and allow them to analyse large amounts of data files in a repeatable manner. Here, we briefly describe the
software, while a detailed step-by-step manual is available at the repository site.\cite{smICA_v2_1_1} The
script is based on the Dear PyGUI - the graphical user interface toolkit (GUI) for Python, making it fast,
dynamic, intuitive, and cross-platform.\\
\input{figure_3}
\indent The smICA script consists of two modes that the user is able to choose at startup. The first mode,
''\emph{EXTRACT from PTU and FILTER}'', is dedicated to
extracting the raw TTTR data from the
binary \ptu\ file that stores the data acquired by the commercial, FLIM/FCS dedicated SymPhoTime software (PicoQuant).
The graphical user interface is shown in Figure \ref{fig:extract_GUI}. The data
are
extracted using the readPTU\_FLIM library\cite{readPTUFLIM}
designed to read the
\ptu\ files. The TTTR data gathered for more than one fluorescence channel (gathered by separate detectors) are split. In each
channel, the user can
filter the data using the time gating method or by applying the statistical filters according to the algorithm
described in references \citenum{enderlein2005using} and \citenum{kapusta2007fluorescence}. As an output, the script returns the
binary pickle file having the form of a dictionary and storing: (\textit{i}) an array consisitng the number of photons
registered in each pixel of the unfiltered image (one array per channel), (\textit{ii}) an array for intensity
values (number of photons after filtration) per pixel (one array per channel), (\textit{iii}) an array for
fluorescence average lifetime values per pixel after filtration (one array per channel), (\textit{iv}) the
decay pattern for the whole considered lifetimes' range, (\textit{v}) the decay pattern for each channel (separate arrays), (\textit{vi}) an array with the filtered background, (\textit{vii}) the TCSPC
pattern used for filtering the signal, (\textit{vii}) the TCSPC pattern for filtering the background, and (\textit{ix}) the TCSPC histogram of the filtered signal.
The software also returns .png files that can be further used to set the
region of interest, ROI, in an external software.\\
\input{figure_4}
\indent The second mode,''\emph{Phot 2 Conc}'', reads in the data created in the ''\emph{EXTRACT from PTU and FILTER}'' mode,
allows for automatically and dynamically defining the ROI for further calculations, and translates the number of photons
registered in each pixel into the
concentration of fluorescent molecules of interest. The user has two options for defining ROI. The first method is based
on the ASCII files created in the
external software; for the purposes of this paper, we used ImageJ. This method, however, requires to use of the
\emph{Rewrite ROI} tool available from the Tools menu. The \emph{Rewrite ROI} tool translates the externally created
files
into files whose structure is recognised by the smICA code. In this method, the user can select the set of ROI files (by
choosing the folder) corresponding to the analyzed
\ptu\ files. There can be more than one ROI file per \ptu\ file, but all ROI files need to be properly named; see manual
\cite{smICA_v2_1_1}.\\
\indent The second option is to use the dynamic detection of ROI based on the Otsu method. In the dynamic
ROI mode,
the Otsu or multi-Otsu thresholding methods are employed to automatically segment the image without requiring manually set
intensity cut-offs. The standard Otsu algorithm evaluates all possible threshold values and selects the one that maximises the
between-class variance (or equivalently minimises the within-class variance) of the pixel intensity histogram. In practice, this
means it assumes a bimodal distribution of pixel values, where one peak corresponds to the background, another to the object and
finds the optimal point separating these two populations. To account for variations in illumination or contrast, the GUI exposes a
ratio parameter that scales the automatically determined threshold, allowing the user to fine-tune segmentation interactively when
Otsu’s estimation slightly under- or over-segments the region.

The multi-OTSU variant generalises this concept by dividing the intensity histogram into three classes, enabling
detection of multiple distinct intensity regions within a single cell. This is particularly useful for identifying
internal cellular structures such as bright or dark nuclei, vesicles, or other subcellular compartments that differ in
fluorescence intensity, depending on the imaging conditions and user input. The ROI created with this method is
automatically saved into the \emph{ROI} subfolder
in case it is required to reuse it. The ROI files created via this method can be further used in the 'file ROI''
mode as they will be automatically recognised by the smICA code.\\
\indent To calculate the concentration of fluorescent molecules per pixel, the software requires the FCS calibration parameters,
including the focal volume structure parameter, $\kappa$,
the lateral size of the focal volume, $\omega$, and the molecular brightness, $B$. All of them can be manually added or
loaded from the \emph{.json} file. The calculations are performed for all files selected by the user, located in the given
folder. As an
output, the script returns the text files containing (\textit{i}) an array of the number of molecules per
pixel, (\textit{ii}) an array of the mean concentration of molecules per pixel, (\textit{iii}) the table in
the user-desired format (\emph{.csv}, \emph{.dat},
\emph{.pickle}\footnote{This is
the binary file storing the pandas DataFrame.}). The output table consists of the File name, ROI number,
channel number, the mean number of counts (photons), mean number of molecules per pixel, mean concentration
per pixel, and the corresponding standard deviations. All mean values correspond to the average over the
entire image or ROI (if selected).

%% file: figure_3.tex
\begin{figure*}[t]
 \includegraphics[width=\linewidth
 ]{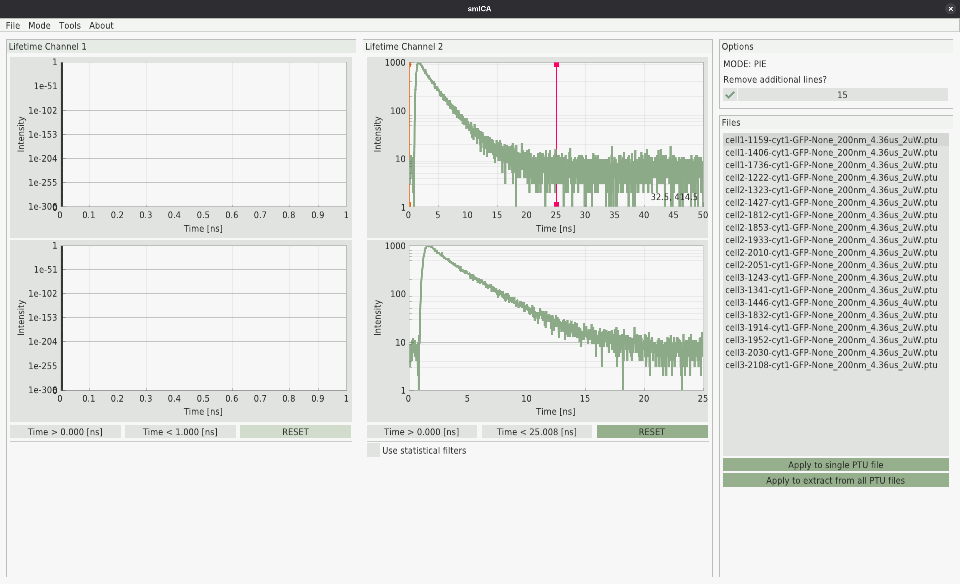}

 \caption{\label{fig:extract_GUI} \textbf{The graphical user interface of the EXTRACT from PTU and FILTER method.} The
central part of the interface takes the TCSPC histograms. The signal is acquired for all pixels and split by detection
channel. The software determines whether the image was acquired in Standard or PIE mode. If the PIE mode is applied, the
time-gating method is used to eliminate crosstalk between fluorescent channels. The selection of the decay-time range is
performed in the upper plots, independently for each channel. The lower plots show the parts of the signal to be
considered in further analysis. The right panel shows the list of files in a given folder. For clarity, the screenshot
was captured in the
\textit{light} theme, while the default theme is \textit{dark} (the user can choose the desired theme). For details, see
the manual on
the repository site.\cite{smICA_v2_1_1}}
\end{figure*}

%% file: figure_4.tex
\begin{figure*}[t]
 \includegraphics[width=\linewidth
 ]{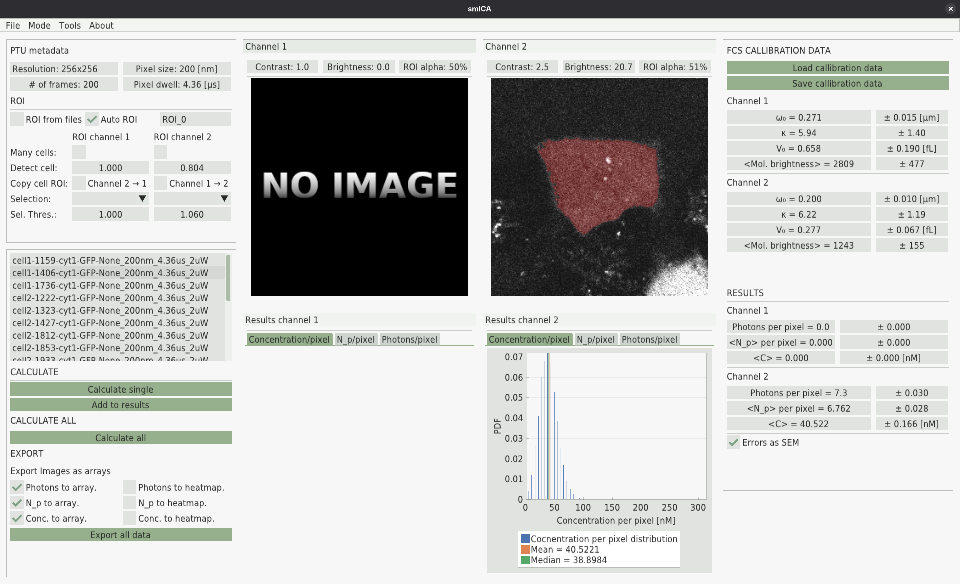}

 \caption{\label{fig:p2c_gui} \textbf{The graphical user interface of the Phot 2 conc method.} The central part of the
interface displays the analysed image (independently for each fluorescent channel). The images show an A549 cell labelled
with an EGFP (channel 2). Below the images are the per-pixel statistics (distribution, mean, and median) for
concentration, the number of fluorophore molecules, and the number of registered photons. On the left side of the panel,
there are: i) the panel displaying the metadata of the selected .ptu file, ii) the ROI control panel, iii) a list of all
data files in a given folder, iv) and the calculate/export control panel. The right-side panels are responsible for
defining calibration data (top) and displaying the mean results (bottom). For clarity, the screenshot was captured in the
\textit{light} theme, while the default theme is \textit{dark} (the user can choose the desired theme). For details, see
the manual on the repository site.\cite{smICA_v2_1_1}}
\end{figure*}

%% file: validation.tex
\subsubsection*{Influence of the experimental variables}
\indent To validate the determination of the concentration, we designed experiments where we changed the fluorophore’s
concentration, the power of the laser $P$ (measured at the sample), the pixel dwell time - $\tpd$, and the size of the pixel $\DeltaX$
or, more precisely, the scan velocity is defined as $\dxdt$.\footnote{The variable controlled experimentally in the hardware setup
is $\DeltaX$. We assume, however, that the effective size of the pixel is equal to the focal volume illuminated within the time
$\tpd$. The focal volume is determined from the FCS calibration measurements.}\\
\indent We measured a well-defined probe, the dextran polymer with a molecular weight of 155 kg/mol, labelled with the TRITC dye.
This polymer is characterised by a hydrodynamic radius of approximately 9 nm. The measurements on dextrans were performed in water
at 25\textcelsius. The concentration of the polymers varied from $\sim$5 to $\sim$100 nM. The laser power, $P$, was changed
between 5 and 40 $\mu$W, which corresponded to the peak focal irradiance, $I_0$,\footnote{$I_0 =
2P\slash\left(\pi\omega_0^2\right)$, where $\omega_0$ stands for the radius of the focal volume.} varying from 0.07 to 0.5
mW/$\mu\mathrm{m}^2$. The pixel dwell time, $\tpd$, was changed between 3.12 $\mu$s and 60 $\mu$s, and the scan velocity changed
between 0.2 nm/$\mu$s and 123 nm/$\mu$s. Another parameter that influences the results is the number of photons registered per
molecule, $\gamma$, which is a product of the molecular brightness, $B$, and $\tpd$. At low laser power (5 $\mu$W, $B$ = 4260
photons/molecule/s) and low pixel dwell (3.12 $\mu$s), the $\gamma$ can be as low as 0.013, which can be understood as the chance
of observing a photon emitted by a single molecule in a given pixel. Even if the $\tpd$ is increased up to 60 $\mu$s and the
laser power is increased to 40 $\mu$W ($B$ = 12300 photons/molecule/s), $\gamma$ is still below unity. To increase the chance of
detecting the photon during image acquisition, we integrated the image over 5 to 100 frames, which allowed us to probe the total
number of photons per molecule, $\Gamma = \gamma\Nf$, between 0.13 and 34, with a median of around 1; here $\Nf$ denotes
the number of
acquired frames.\\
\input{figure_5}
\indent Based on the above experimental variables, we determined four parameters that primarily influence the concentration value
determined by our method: the pixel size ($\DeltaX$), the total number of photons registered per molecule ($\Gamma$), molecular
brightness ($B$), and scan velocity ($\dxdt$). We performed $\sim$460 measurements, varying those parameters, and
conducted a statistical analysis of the results. We assumed that the concentration in the given measurement is correctly
determined if the $\ratC$ ratio, being the ratio of the concentration determined from smICA to the concentration determined from
FCS, falls within the interval 0.75-1.25. Next, we calculated the two-dimensional maps of
the probability of success $\hat{P}\left(x,y\right)$ where $x,y$ is one of the pairs ($\DeltaX$, $\Gamma$) or
($\dxdt$, $B$). The probability surface was estimated using the non-parametric Nadaraya–Watson kernel
estimator\cite{Watson1964, Nadaraya1964} with the Gaussian kernel, according to the following procedure. First, we standardized
the experimental observations for each pair using robust z-scores (median/MAD; MAD - median
absolute deviation) to ensure comparable scaling. At each point ($x_0$, $y_0$) of the regular grid, we adaptively choose the
bandwidth $h\left(x_0,y_0\right)$ using the Euclidean distance to the $k$-th nearest neighbour \cite{Demir2010} in the training
data (we set $k=40$). Next, we estimated the conditional success probability as:
\begin{equation}
 \hat{P}(x_0, y_0) =
\frac{
    \displaystyle\sum_{i=1}^{n}
    K_{h(x_0, y_0)}\!\left( \left\| (x_0, y_0) - (x_i, y_i) \right\| \right) \, y_i
}{
    \displaystyle\sum_{i=1}^{n}
    K_{h(x_0, y_0)}\!\left( \left\| (x_0, y_0) - (x_i, y_i) \right\| \right)
}
\end{equation}
where $y_i$ is the indicator function (1 - the $i$-th measurement was successful, 0 otherwise) and $K_h\left(r\right) =
\exp\left[-\left(r\slash h\right)^2\slash2\right]$ is the Gaussian kernel with bandwidth $h$. The effective local sample size was
computed as:
\begin{equation}
 \neff(x_0, y_0)
= \frac{\left( \sum_i w_i \right)^2}{\sum_i w_i^2},
\qquad
w_i = K_{h(x_0, y_0)}\!\left( \left\| (x_0, y_0) - (x_i, y_i) \right\| \right)
\end{equation}
Regions with insufficient data support ($\neff<100$) were shaded with inclined lines. The probability maps are shown in Figure
\ref{fig:maps}.
For a broad range of experimental parameters, the chance that the smICA method correctly measures the concentration of the
fluorophore is higher than  80\%. The lowest probability is estimated when the pixel size is too small ($\DeltaX$ < 100 nm) and
for dim molecules. Increasing the number of frames used to create the image ($\Nf$) raises $\Gamma$, and the
$\hat{P}$. For example, $\hat{P}$ increases above 93\% for $\Gamma\sim 8$ $\frac{\mathrm{Photons}}{\mathrm{Molecule}}$. For the
range of $\tpd$ used in this study, $\DeltaX <
100$
resulted in a
small scan velocity values ($\dxdt \lesssim35$). In this region, when the molecule is dim ($B\lesssim5\times10^3$
$\frac{\mathrm{Photons}}{\mathrm{Molecule}\cdot s}$ low laser power or quantum yield of fluorophore), the $\hat{P}$ was estimated
at $\lesssim70\%$. Increase of the scan velocity or the molecular brightness of the molecule increases $\hat{P}$  above 93\% (for
$B\approx10^4$ $\frac{\mathrm{Photons}}{\mathrm{Molecule}\cdot s}$).

\subsubsection*{Influence of ROI size and filtration}
\indent Quantitative microscopy techniques are primarily developed for biological applications to monitor intracellular processes.
Those types of measurements, however, often require selective analysis of specific regions in cells or tissues and are prone to
artefacts, such as autofluorescence or dye-dye crosstalk. We challenged our methodology against those factors. One of
the methods that allows for filtering unwanted photons is to remove them directly from the TTTR signal before the
over-all-frames
integration. To achieve this, we employed a method widely used in fluorescence correlation spectroscopy, which is based on
fluorescence lifetime decay phenomena. Below, we compare and discuss the TTTR signal filtering using measurements performed
in a simple \textit{in vitro} system and in living cells.\\
\input{figure_6}

%
%
\indent We used the raw experimental data, which we previously presented in Figure
\ref{fig:maps} for aqueous dextran solutions. We selected a series of different-sized regions of interest with the
number of
pixels, $\npx$, varying from 16$^2$ to 256$^2$ pixels ($\npx$=256$^2$ refers to the full frame). For each $\npx$ value (except
$\npx=256^2$) in every image, we selected seven randomly positioned regions of interest. The concentration was calculated for each
of these ROIs and then averaged. This procedure allowed us to
verify the influence of ROI size on the $\ratC$ ratio. The procedure was repeated using two methods of pattern filtering. In the
first method, we used the fluorescence decay pattern of the fluorophore calculated
automatically by subtracting the background noise level (average number of counts in the TCSPC histogram for decay times
exceeding 15 ns; compare the Non-filtered signal and the TCSCP-Background free signal in Figure
\ref{fig:fitered_unfiltered}\textbf{a})
from the raw TCSPC histogram. In the second method, we used a
predefined, high-quality fluorescence decay pattern registered for the same fluorophore in the same system, but at higher
concentration and laser power, and appplied the TCSPC pattern filtration\cite{enderlein2005using,
kapusta2007fluorescence} of the raw TTTR
signal, which resulted in the smooth but less intense signal marked as the TCSPC-Pattern filtered signal in Figure
\ref{fig:fitered_unfiltered}\textbf{a}. In both cases, we analysed the $\ratC$ ratio.\\
\indent In the experiments used for this test, we have only one fluorophore, and we do not expect any artefacts that could
negatively affect the concentration values. In such a case, one can expect that the filtration of the signal using the
fluorescence decay pattern may overfilter the data. Indeed, diagrams depicting the statistics of both $\ratC$ ratios for various
ROI sizes, presented in Figures \ref{fig:fitered_unfiltered}\textbf{b}, exhibit strong deviations toward lower
$\ratC$ values, indicating an underestimation of $\Cim$. The size of the ROI does not influence the overall results, though the
distribution of data becomes narrower with increasing $\npx$.
\input{figure_7}
\indent The \textit{in vitro} studies described above on dextrans demonstrate that, in such a simple system, sophisticated
filtration of the TTTR signal is not required and can be detrimental to the analysis. In the case of biological samples, however, we
can expect
more difficulties to overcome. In such cases, we can expect the influence of autofluorescence and/or dye-dye crosstalk, exhibited
as the leaking of photons between fluorescence channels and resulting from overlapping fluorescence spectra. One method
of filtering the signal is time-gating.\cite{Rich2013} In this method, the signal is taken into consideration only if the
photons arrive in a certain range of time after the laser pulse. The second method is the same as the one we employed to analyse
the \textit{in vitro} data and is based on the TCSPC pattern. We applied both methods and compared their results to the raw,
non-filtered
data.\\
\indent We performed test measurements on the cells where
two types of fluorophores were present: the EGFP and the TRITC-labelled dextran. The TRITC-labelled dextran was mixed with the
non-labelled mRNA at known proportions and injected into cells. The EGFP was synthesised in cells after injection
of the mRNA encoding the EGFP sequence. To separate the excitation events between the
laser lines, we operated in pulse interleaved excitation (PIE) mode, allowing us to shift the consecutive pulses of the lasers in
time. We used a shift between consecutive pulses equal to 25 ns, while the delay time between pulses of the same laser (same
wavelength) was 50 ns. We analysed data from 17 different cells. The power of the lasers was set to 10 $\mu$W for the green laser
($\lambda$ = 561 nm) and 2 $\mu$W for the blue laser ($\lambda$ = 485 nm). The images were acquired for $\DeltaX$ = 399 nm,
$\tpd$ = 8 $\mu$s and integrated over 100 frames. \\
\indent Figures \ref{fig:fitered_unfiltered_gfp_dex}\textbf{a} and \textbf{b} show representative fluorescence decay patterns of
the
TRITC-labelled dextran and the EGFP, acquired with two SPAD detectors corresponding to the
TRITC and EGFP channels. The pulse of the green laser is shifted by +25 ns with respect to the pulse of the blue laser. In the
time-gating filtration method, we limit the range of decay times for each channel. For the TRITC channel, we selected a decay-time
range of 25-50 ns (Figure \ref{fig:fitered_unfiltered_gfp_dex}\textbf{a}). Consequently, the range of decay times for the EGFP
channel was chosen between 0 and 25 ns; Figure \ref{fig:fitered_unfiltered_gfp_dex}\textbf{b}. In the TCSPC filtration
method, apart
from time gating, we applied background removal in the same manner as depicted in Figure \ref{fig:fitered_unfiltered}\textbf{a}.
Figure \ref{fig:fitered_unfiltered_gfp_dex}\textbf{c} compares the $\ratC$ ratio for the non-filtered and filtered
signals for
TRITC-labelled dextran and EGFP, respectively. The non-filtered signal exhibits a slight overestimation of the concentration for
both channels and most measurements. Both the TCSPC-based method of filtration (time gating and TCSPC background removal
filtration) display $\ratC$ values close to unity.\\

%% file: figure_5.tex
\begin{figure*}[t]
 \includegraphics[width=\linewidth
 ]{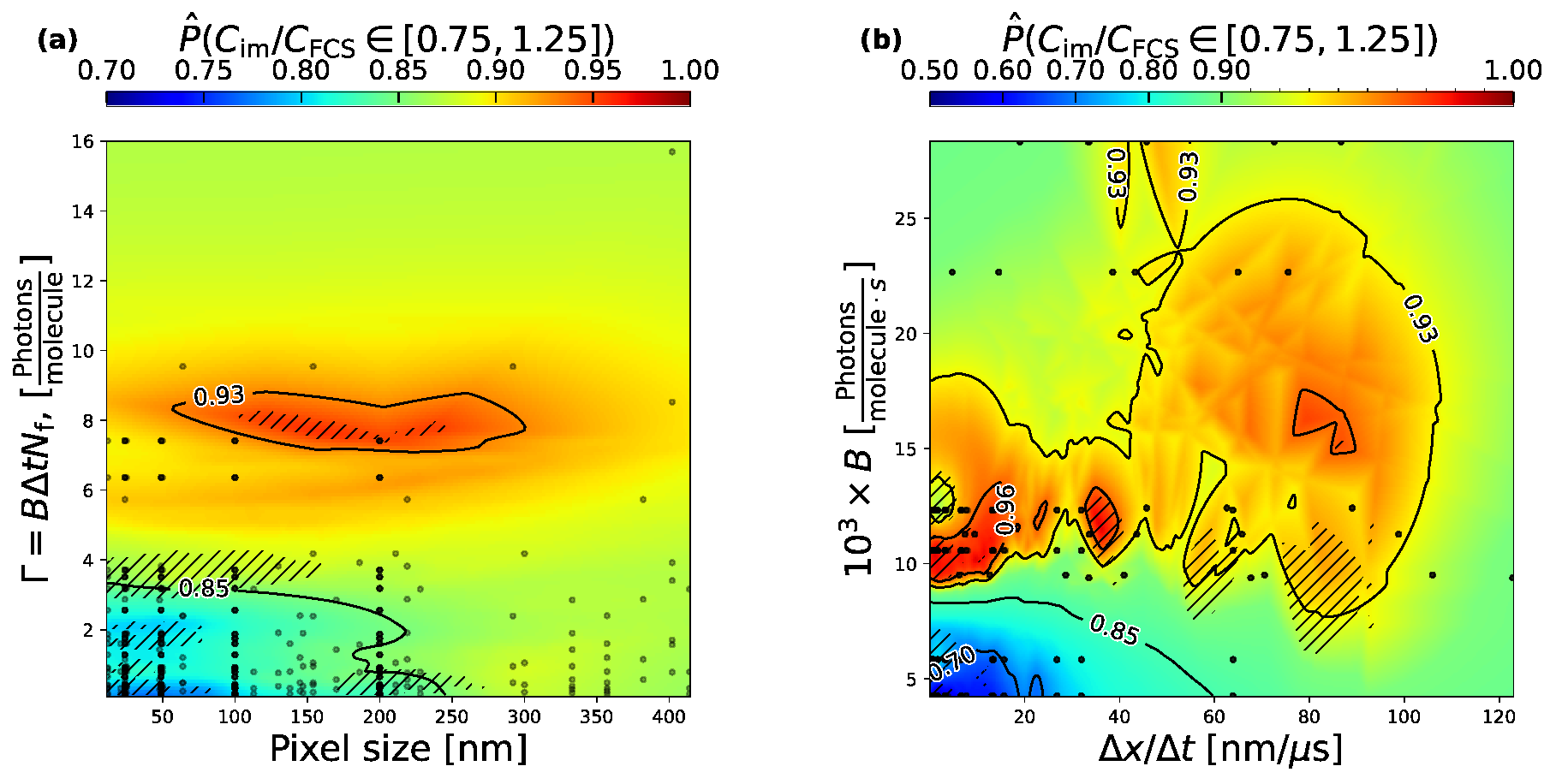}

 \caption{\label{fig:maps} \textbf{Statistical analysis of the influence of experimental parameters on the determination of concentration.} Panels depict heatmaps of the estimated conditional probability $\hat{P}$ that the $\ratC$ takes values between 0.75 and 1.25. $\hat{P}$ was estimated from the experimental data of $\ratC$ for various pairs of parameters: \textbf{(a)} pixel size \vs. total expected number of
photons per molecule ($\Gamma$), \textbf{(b)} scan velocity $\dxdt$ \vs. molecular brightness, $B$. The $\hat{P}$ was estimated
utilising the Nadaraya–Watson method\cite{Watson1964, Nadaraya1964} using the isotropic Gaussian kernel with adaptive bandwidth
being the $k$-th ($k$=40) nearest neighbour distance.\cite{Demir2010} Black dots marks the experimental points, solid lines
indicate $\hat{P}$ contour lines. The regions marked with /// depict regions of low trust where the effective local sample size,
$\neff<100$.}
\end{figure*}

%% file: figure_6.tex
\begin{figure}[t]
 \includegraphics[width=\linewidth
 ]{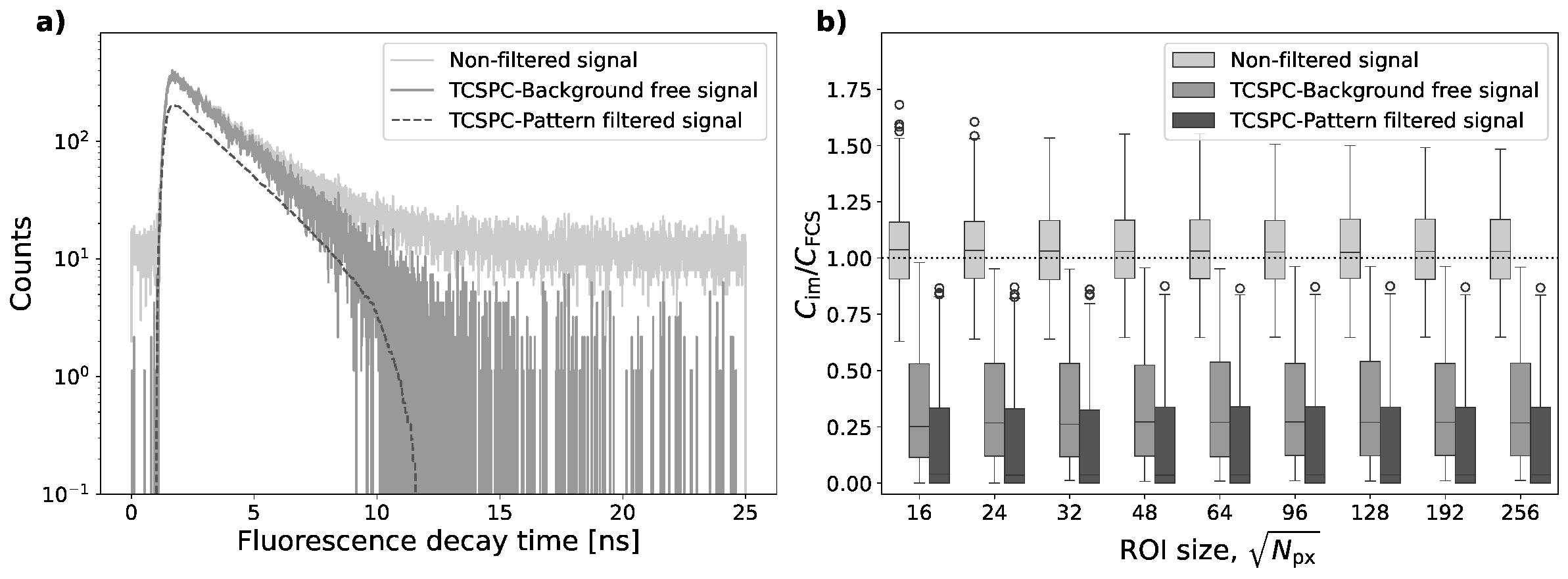}

 \caption{\label{fig:fitered_unfiltered}\textbf{The influence of filtration and Region of interest size on the imaging of
concentration.} Panel \textbf{a} depicts a plot consisting of the fluorescence decay patterns for the non-filtered signal, the
signal filtered by the TCSPC background removal, and the TCSPC pattern filtration methods; all acquired for the TRITC-labelled
dextran polymer in water. Both filtering methods were based on the algorithm described in references \citenum{enderlein2005using},
and \citenum{kapusta2007fluorescence}. In the background removal method, we subtracted the constant value of the background (see
main text) to find the TCSPC pattern to be filtered. In the TCSPC-pattern method, we applied a previously registered pattern of
fluorescence decay acquired in a system with a sufficiently high S/N ratio. Panel \textbf{b} shows box plots representing the
dispersion of the $\ratC$ ratio data obtained for different region-of-interest sizes. ROI size $\sqrt{\npx} = 256$ corresponds to
the full frame.}
\end{figure}

%% file: figure_7.tex
\begin{figure}[t!]
 \includegraphics[width=0.9\linewidth
 ]{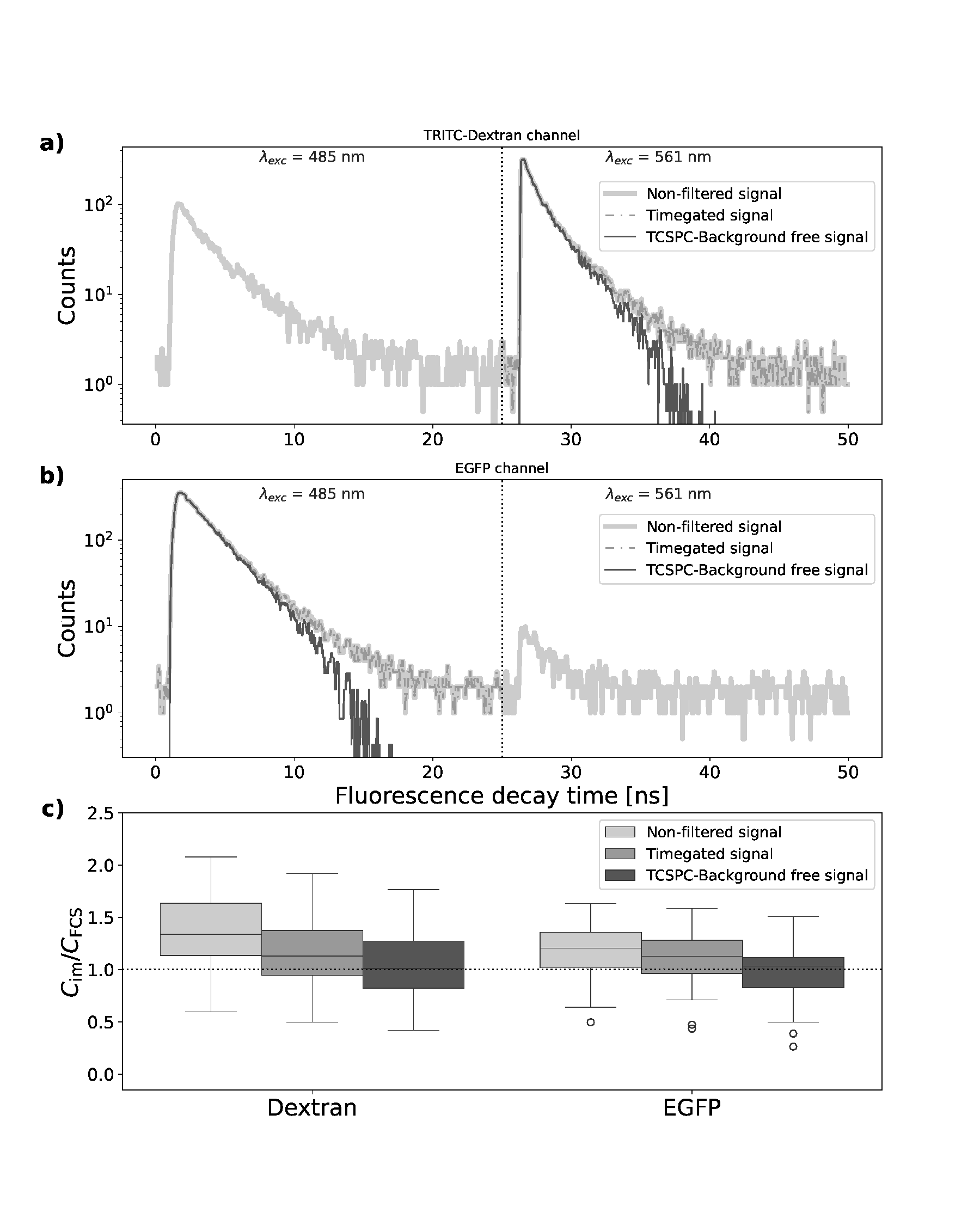}
 \caption{\label{fig:fitered_unfiltered_gfp_dex} \textbf{Application of the TCSPC-based filtration to the imaging of concentration
in living cells.} Figures \textbf{a} and \textbf{b} depict raw fluorescence decay patterns acquired in the pulse interleaved
excitation, PIE, mode in living cells where two fluorophores (EGFP and TRITC-labelled dextran) were present. Figure \textbf{a}
shows the data registered in the TRITC-channel ($\lambda_{\mathrm{em}}>594$ nm). Although the photons were registered in the TRITC
channel that was hardware-filtered by the beamsplitter, some EGFP molecules emitted the photons of wavelengths exceeding the
$\lambda_{\mathrm{em}}>594$ nm, due to the long tail of the emission spectra of the fluorescent protein. By analogy, Figure
\textbf{b} depicts the fluorescence decay for the photons registered in the EGFP channel ($\lambda_{\mathrm{em}} = 525 \ pm 25$
nm, the band-pass filter). Here, part of the TRITC molecules were also excited by
the 485 nm laser. Figure \textbf{c} shows boxplots comparing the $\ratC$ ratio for the filtered and non-filtered data. The
filtering was performed using a simple time-gating method, which involved selecting photons that fell within the specified
decay-time range, along with the TCSPC background removal method.
}
\end{figure}

%% file: Experimental_example.tex
\input{figure_8}
\indent To visualise potential application of our methodology and capabilities of the smICA software, we performed exemplary
measurements of the decay of exogenous mRNA concentration in living cells. The exemplary A549 cell was injected with the
sulfo-Cy5-labelled mRNA. After the injection, we monitored the same cell for 10 hours. We observed that, over time, the mRNA
molecules are enclosed in globular structures approximately 1 $\mum$ in diameter. We monitored the concentration in three regions
of interest: \textit{i)} in the cytoplasm of the cell (ROI excluded the nucleus), \textit{ii)} the cytosol (ROI excluded the
nucleus and bright spots), \textit{iii)} the bright granular spots only. For each image representing a single time point, we used
smICA's automatic ROI functionality: first selecting the whole-cell ROI, then the find dark option to select the nucleus, and
finally the find many bright spots option to detect granular structures. Next, using the smICA ROI mixer tool, we subtracted
subsequent ROIs to obtain the desired pattern. Figure \ref{fig:exp_cells}\textbf{a} depicts the raw intensity image of the cell
and the ROI patterns used for each time point of the experiment. Finally, using the ROIs from Figure
\ref{fig:exp_cells}\textbf{a}, we calculated changes in the concentration of fluorescently labelled mRNA within each ROI. The
results of the concentration decay are plotted in Figure \ref{fig:exp_cells}\textbf{b}.

%% file: figure_8.tex
\begin{figure}[ht]
 \includegraphics[width=0.95\linewidth
 ]{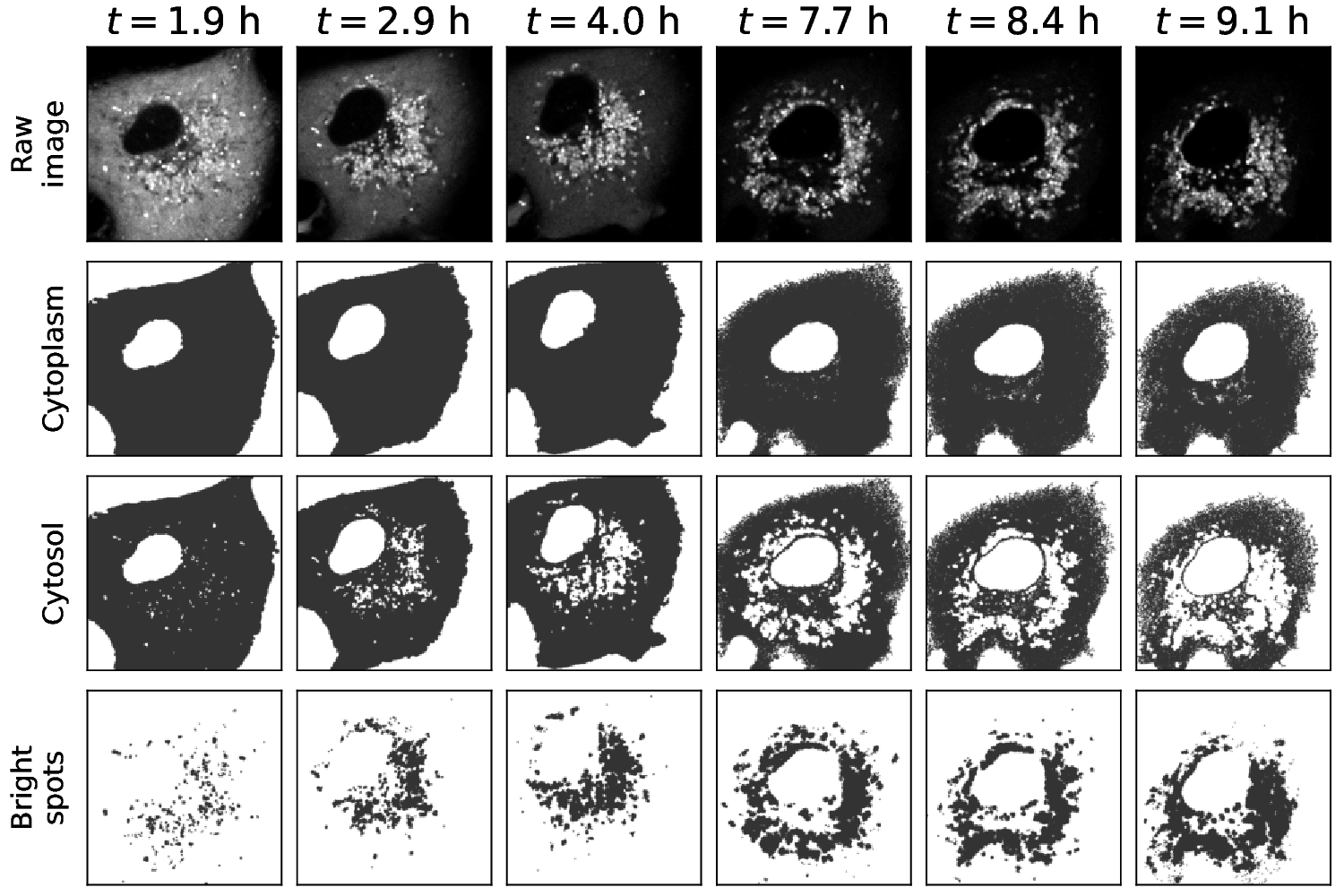}\\
 \includegraphics[width=0.95\linewidth
 ]{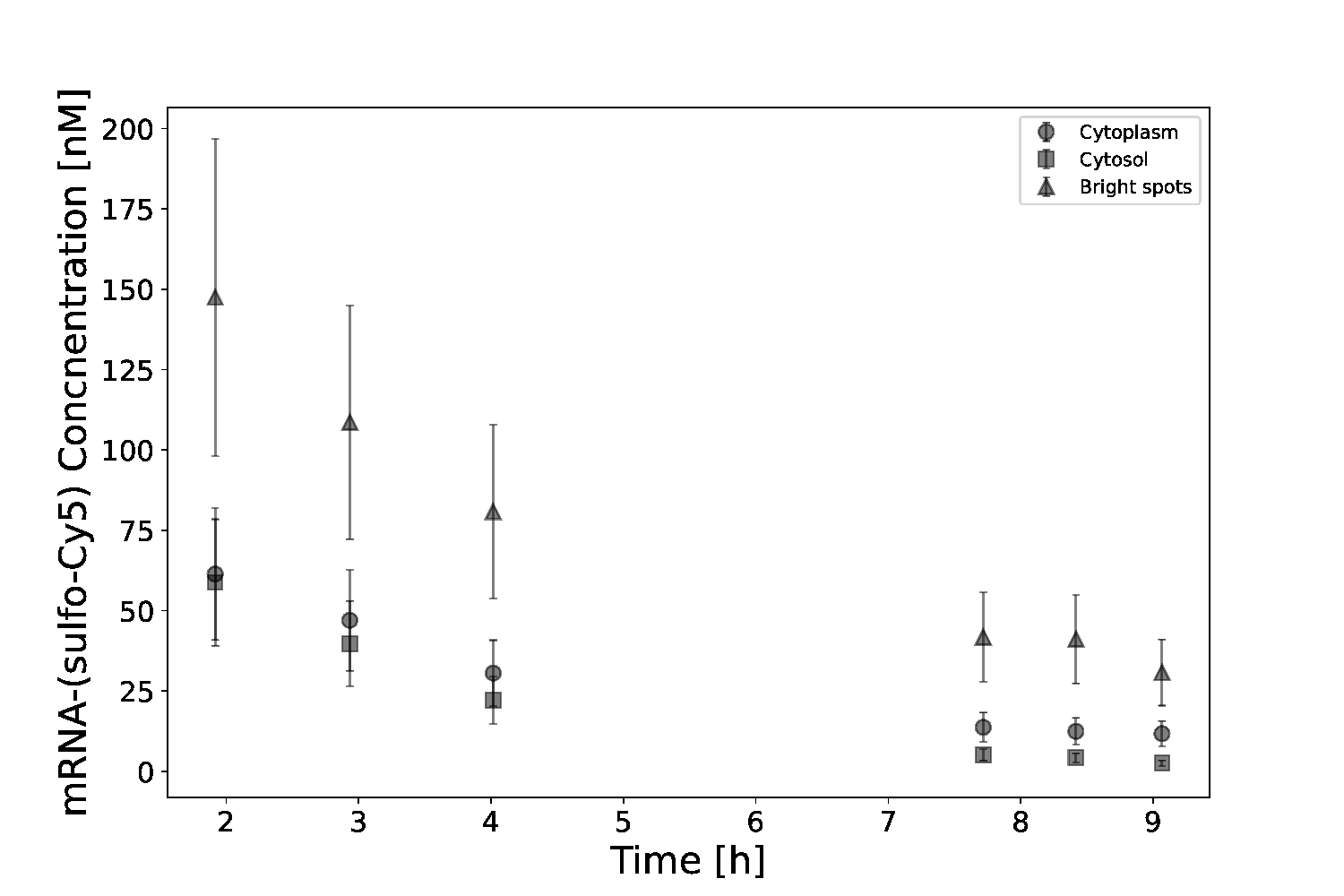}
 \caption{\label{fig:exp_cells}\textbf{Exemplary application of the smICA software for studying the mapping of changes in mRNA
concentration in living cells.} Figure \textbf{a)} shows the time series of images for single cells (top row), together with the
ROI used to map mRNA concentration. The black region in the ROIs panel represents the regions of interest within cells used to
calculate mRNA concentration. Figure \textbf{b)} depicts the plot of the mean concentration calculated from the corresponding
ROI. Every single point represents the multiframe image (200 frames) acquired with the pixel dwell time of 4.36 $\mu\mathrm{s}$.
The power of the 641 nm laser at the sample plane was 4 $\mu\mathrm{W}$. The pixel dwell time, the laser power and the time
between consecutive measurements (around an hour) were chosen to minimise the photobleaching of the sample. The error bars
represent the mean per-pixel propagated uncertainty within the region of interest (ROI), derived from the focal-volume and
molecular-brightness uncertainties obtained during calibration.}
\end{figure}

%% file: Conclusions.tex
We presented the smICA - a methodology and the GUI software for mapping the concentration of fluorophores in living cells. The
method requires calibration of the confocal volume using standard FCS measurements and assumes that the brightness of the
fluorophore is constant in time and space, which is reasonable as long as the fluorophore's local environment (mostly water) does
not change. Meeting both requirements allows us to determine the mean concentration per pixel, averaged over the data
acquisition time. Therefore, an output contains information about the average concentration over the entire cell/ROI and the
concentration map across the region.\\
\indent We validated the method by performing the measurements in aqueous solutions of
fluorescently labelled polymers as well as in living cells. The robustness of the method was confirmed by tests spanning
a broad range of
experimental parameters, such as the molecular brightness of the fluorophore, pixel size, pixel dwell time, and the size of the
analysed image. Using the non-parametric statistical analysis with the Nadaraya–Watson kernel estimator, we showed that for
most settings, the probability $\hat{P}$ that the method determines the concentration with an error of < 25\% is higher than
85\%. $\hat{P}$ drops below 70\%  only when the pixel sizes $\DeltaX<$ 100 nm and molecular brightness $\sim$ 5000
Photons/molecule/s, simultaneously. Even then, the probability $\hat{P}$ can be increased to $\hat{P}>85\%$ by integrating
over tens to hundreds of frames. We demonstrated that for a system composed of a single type fluorophore (TRITC-dextran
in
water), the signal does not require additional filtering. In the case of cellular measurements, performed using EGFP and
TRITC-dextran as the probes, it is recommended to filter the signal using fluorescence decay time
pattern filtering, either by
simple time gating or by subtracting the background, as it is performed in fluorescence lifetime correlation
spectroscopy (FLCS).\\
\indent Currently, the methodology and software developed for these purposes are limited to the data acquired using the
TCSPC setup of a
single vendor. Nevertheless, the software is provided as an open-source project and can be easily expanded to other
LSCMs and
FCS vendors. Moreover, our results suggest that the presented methodology, along with the software, is a promising tool
for
quantitative single-cell studies.\\
\indent The smICA software has the unique feature of mapping absolute concentrations, which we illustrated in the example of
mapping the time evolution of exogenous mRNA concentration in a living cell. The experiment shows that smICA is a promising
bioimaging tool with potential applications in studies focused on tracing the concentration of the probe of interest within the
cell and its sub-compartments. The filtering of the image data with fluorescence decay time patterns, implemented in our software,
provides insights into more detailed topics. These topics include protein expression, biomolecule degradation, monitoring
enzymatic reactions, and investigating the affinity of probes for specific cellular targets.

%% file: PTU_CONC_references.bib
@Article{Vukojevic2008,
  author       = {Vukojevi{\'c}, Vladana and Heidkamp, Marcus and Ming, Yu and Johansson, Bj{\"o}rn and Terenius, Lars and Rigler, Rudolf},
  journal      = {Proceedings of the National Academy of Sciences},
  title        = {Quantitative single-molecule imaging by confocal laser scanning microscopy},
  year         = {2008},
  number       = {47},
  pages        = {18176--18181},
  volume       = {105},
  creationdate = {2023-11-07T16:41:40},
  publisher    = {National Acad Sciences},
}

@article{Digman2008,
  title={Mapping the number of molecules and brightness in the laser scanning microscope},
  author={Digman, Michelle A and Dalal, Rooshin and Horwitz, Alan F and Gratton, Enrico},
  journal={Biophysical journal},
  volume={94},
  number={6},
  pages={2320--2332},
  year={2008},
  publisher={Elsevier}
}

@Article{Magde1972,
  author    = {Magde, Douglas and Elson, Elliot and Webb, Watt W},
  journal   = {Physical review letters},
  title     = {Thermodynamic fluctuations in a reacting system—measurement by fluorescence correlation spectroscopy},
  year      = {1972},
  number    = {11},
  pages     = {705},
  volume    = {29},
  publisher = {APS},
}

@Article{Elson1974,
  author    = {Elson, Elliot L and Magde, Douglas},
  journal   = {Biopolymers: Original Research on Biomolecules},
  title     = {Fluorescence correlation spectroscopy. I. Conceptual basis and theory},
  year      = {1974},
  number    = {1},
  pages     = {1--27},
  volume    = {13},
  publisher = {Wiley Online Library},
}

@Article{Ehrenberg1974,
  author    = {Ehrenberg, M{\aa}ns and Rigler, Rudolph},
  journal   = {Chemical Physics},
  title     = {Rotational brownian motion and fluorescence intensify fluctuations},
  year      = {1974},
  number    = {3},
  pages     = {390--401},
  volume    = {4},
  publisher = {Elsevier},
}

@Article{Koppel1974,
  author    = {Koppel, Dennis E},
  journal   = {Physical Review A},
  title     = {Statistical accuracy in fluorescence correlation spectroscopy},
  year      = {1974},
  number    = {6},
  pages     = {1938},
  volume    = {10},
  publisher = {APS},
}

@Article{Brown2008,
  author         = {Brown, CM and Dalal, RB and Hebert, B and Digman, MA and Horwitz, AR and Gratton, E},
  journal        = {Journal of microscopy},
  title          = {Raster image correlation spectroscopy (RICS) for measuring fast protein dynamics and concentrations with a commercial laser scanning confocal microscope},
  year           = {2008},
  number         = {1},
  pages          = {78--91},
  volume         = {229},
  publisher      = {Wiley Online Library},
  qualityassured = {qualityAssured},
}

@article{mamot2022ethylenediamine,
  title={Ethylenediamine derivatives efficiently react with oxidized RNA 3' ends providing access to mono and
dually labelled RNA probes for enzymatic assays and in vivo translation},
  author={Mamot, Adam and Sikorski, Pawel J and Siekierska, Aleksandra and de Witte, Peter and Kowalska,
Joanna and Jemielity, Jacek},
  journal={Nucleic Acids Research},
  volume={50},
  number={1},
  pages={e3--e3},
  year={2022},
  publisher={Oxford University Press}
}

@article{kapusta2007fluorescence,
  title={Fluorescence lifetime correlation spectroscopy},
  author={Kapusta, Peter and Wahl, Michael and Benda, Ale{\v{s}} and Hof, Martin and Enderlein, J{\"o}rg},
  journal={Journal of fluorescence},
  volume={17},
  pages={43--48},
  year={2007},
  publisher={Springer}
}

@article{enderlein2005using,
  title={Using fluorescence lifetime for discriminating detector afterpulsing in fluorescence-correlation
spectroscopy},
  author={Enderlein, J{\"o}rg and Gregor, Ingo},
  journal={Review of scientific instruments},
  volume={76},
  number={3},
  year={2005},
  publisher={AIP Publishing}
}

@Article{Wahl2014,
  author  = {Wahl, Michael and Orthaus-M{\"u}ller, Sandra},
  journal = {Technical Note. PicoQuant GmbH, Germany},
  title   = {Time tagged time-resolved fluorescence data collection in life sciences},
  year    = {2014},
  pages   = {1--10},
  volume  = {2},
}

@Article{Kapusta2010,
  author  = {Kapusta, Peter},
  journal = {Application Note. PicoQuant GmbH, Germany},
  title   = {Absolute Diffusion Coefficients: Compilation of Reference Data for FCS Calibration},
  year    = {2010},
  pages   = {1--2},
}

@article{Ohsugi2009,
  title={Multipoint fluorescence correlation spectroscopy with total internal reflection fluorescence microscope},
  author={Ohsugi, Yu and Kinjo, Masataka},
  journal={Journal of Biomedical Optics},
  volume={14},
  number={1},
  pages={014030--014030},
  year={2009},
  publisher={Society of Photo-Optical Instrumentation Engineers}
}

@article{Galland2011,
  title={Multi-confocal fluorescence correlation spectroscopy: experimental demonstration and potential applications for living
cell measurements},
  author={Galland, R{\'e}mi and Gao, Jie and Kloster, Meike and Herbomel, Gaetan and Destaing, Olivier and Balland, Martial and
Souchier, Catherine and Usson, Yves and Derouard, Jacques and Wang, Ir{\`e}ne and others},
  journal={arXiv preprint arXiv:1112.5392},
  year={2011}
}

@article{Yamamoto2018,
  title={Multipoint fluorescence correlation spectroscopy using spatial light modulator},
  author={Yamamoto, Johtaro and Mikuni, Shintaro and Kinjo, Masataka},
  journal={Biomedical Optics Express},
  volume={9},
  number={12},
  pages={5881--5890},
  year={2018},
  publisher={Optica Publishing Group}
}

@article{Sugiyama2005,
  title={Determination of absolute protein numbers in single synapses by a GFP-based calibration technique},
  author={Sugiyama, Yoshiko and Kawabata, Izumi and Sobue, Kenji and Okabe, Shigeo},
  journal={Nature methods},
  volume={2},
  number={9},
  pages={677--684},
  year={2005},
  publisher={Nature Publishing Group US New York}
}

@Article{Watson1964,
  author    = {Watson, Geoffrey S},
  journal   = {Sankhy{\=a}: The Indian Journal of Statistics, Series A},
  title     = {Smooth regression analysis},
  year      = {1964},
  pages     = {359--372},
  publisher = {JSTOR},
}

@Article{Nadaraya1964,
  author    = {Nadaraya, Elizbar A},
  journal   = {Theory of Probability \& Its Applications},
  title     = {On estimating regression},
  year      = {1964},
  number    = {1},
  pages     = {141--142},
  volume    = {9},
  publisher = {SIAM},
}

@Article{Demir2010,
  author    = {Demir, Serdar and Toktami{\c{s}}, {\"O}niz},
  journal   = {Hacettepe Journal of Mathematics and Statistics},
  title     = {On the adaptive Nadaraya-Watson kernel regression estimators},
  year      = {2010},
  number    = {3},
  pages     = {429--437},
  volume    = {39},
  publisher = {Hacettepe University},
}

@Article{Aguilera2025,
  author       = {Aguilera, Luis U. and Raymond, William S. and Sears, Rhiannon M. and Nowling, Nathan L. and Munsky, Brian and Zhao, Ning},
  journal      = {bioRxiv},
  title        = {MicroLive: An Image Processing Toolkit for Quantifying Live-cell Single-Molecule Microscopy},
  year         = {2025},
  doi          = {10.1101/2025.09.25.678587},
  elocation-id = {2025.09.25.678587},
  eprint       = {https://www.biorxiv.org/content/early/2025/09/29/2025.09.25.678587.full.pdf},
  publisher    = {Cold Spring Harbor Laboratory},
  url          = {https://www.biorxiv.org/content/early/2025/09/29/2025.09.25.678587},
}

@Article{Rich2013,
  author    = {Rich, Ryan M and Stankowska, Dorota L and Maliwal, Badri P and S{\o}rensen, Thomas Just and Laursen, Bo W and Krishnamoorthy, Raghu R and Gryczynski, Zygmunt and Borejdo, Julian and Gryczynski, Ignacy and Fudala, Rafal},
  journal   = {Analytical and bioanalytical chemistry},
  title     = {Elimination of autofluorescence background from fluorescence tissue images by use of time-gated detection and the AzaDiOxaTriAngulenium (ADOTA) fluorophore},
  year      = {2013},
  number    = {6},
  pages     = {2065--2075},
  volume    = {405},
  publisher = {Springer},
}

@Article{Politi2018,
  author    = {Politi, Antonio Z and Cai, Yin and Walther, Nike and Hossain, M Julius and Koch, Birgit and Wachsmuth, Malte and Ellenberg, Jan},
  journal   = {Nature protocols},
  title     = {Quantitative mapping of fluorescently tagged cellular proteins using FCS-calibrated four-dimensional imaging},
  year      = {2018},
  number    = {6},
  pages     = {1445--1464},
  volume    = {13},
  publisher = {Nature Publishing Group},
}

@Article{Brinkenfeldt2023,
  author    = {Brinkenfeldt, Nikolaj K and Dias, Andr{\'e} and Moreno-Pescador, Guillermo and Bendix, Poul Martin and Martinez, Karen L},
  journal   = {bioRxiv},
  title     = {Quantitative Determination of Protein Concentrations in Living Cells},
  year      = {2023},
  pages     = {2023--05},
  publisher = {Cold Spring Harbor Laboratory},
}

@Article{Lo2015,
  author    = {Lo, Chiu-An and Kays, Ibrahim and Emran, Farida and Lin, Tsung-Jung and Cvetkovska, Vedrana and Chen, Brian Edwin},
  journal   = {Cell reports},
  title     = {Quantification of protein levels in single living cells},
  year      = {2015},
  number    = {11},
  pages     = {2634--2644},
  volume    = {13},
  publisher = {Elsevier},
}

@Article{Pilz2025,
  author    = {Pilz, Marta and Kalwarczyk, Tomasz and Burdzy, Krzysztof and Ho{\l}yst, Robert},
  journal   = {Talanta},
  title     = {Quantitative analysis of transferrin uptake into living cells at single-molecule level},
  year      = {2025},
  pages     = {127031},
  volume    = {282},
  publisher = {Elsevier},
}

@Article{Ranjit2018,
  author    = {Ranjit, Suman and Malacrida, Leonel and Jameson, David M and Gratton, Enrico},
  journal   = {Nature protocols},
  title     = {Fit-free analysis of fluorescence lifetime imaging data using the phasor approach},
  year      = {2018},
  number    = {9},
  pages     = {1979--2004},
  volume    = {13},
  publisher = {Nature Publishing Group UK London},
}

@Article{Digman2008a,
  author    = {Digman, Michelle A and Caiolfa, Valeria R and Zamai, Moreno and Gratton, Enrico},
  journal   = {Biophysical journal},
  title     = {The phasor approach to fluorescence lifetime imaging analysis},
  year      = {2008},
  number    = {2},
  pages     = {L14--L16},
  volume    = {94},
  publisher = {Elsevier},
}

@software{smICA_v2_1_1,
  author       = {Kalwarczyk, Tomasz},
  title        = {smICA - single-molecule Image to Concentration
                   Analyser
                  },
  month        = may,
  year         = 2026,
  publisher    = {Zenodo},
  version      = {v2.1.1},
  doi          = {10.5281/zenodo.20134967},
  url          = {https://doi.org/10.5281/zenodo.20134967},
  note      = {Version 2.1.1, Zenodo, Software, DOI:10.5281/zenodo.20134967. (The development repository is hosted on
GitHub: \url{https://tkmist.github.io/smICA/}.)},
}

@Misc{readPTUFLIM,
  author = {Rohilla, Sumeet},
  note   = {The development repository is hosted on GitHub: \url{https://github.com/SumeetRohilla/readPTU\_FLIM}. For the puroses of the smICA software we used modified version of the repository from the branch entitled "NIKON\_correction" available at \url{https://github.com/TKmist/readPTU\_FLIM/NIKON\_correction}},
  title  = {readPTU\_FLIM},
  year   = {2019},
}

@misc{FcsIT_Kalwarczyk2026_arXiv,
      title={FcsIT: An Open-Source, Cross-Platform Tool for Correlation and Analysis of Fluorescence Correlation
Spectroscopy Data},
      author={Tomasz Kalwarczyk},
      year={2026},
      eprint={2603.29684},
      archivePrefix={arXiv},
      primaryClass={q-bio.QM},
      url={https://arxiv.org/abs/2603.29684},
}

@software{kalwarczyk_2026_19367803,
  author       = {Kalwarczyk, Tomasz},
  title        = {FcsIT - A simple and easy-to-use tool for
                   correlating and fitting the fluorescence
                   correlation spectroscopy (FCS) data.
                  },
  month        = apr,
  year         = 2026,
  publisher    = {Zenodo},
  version      = {v1.0.2},
  doi          = {10.5281/zenodo.19367803},
  url          = {https://doi.org/10.5281/zenodo.19367803},
  swhid        = {swh:1:dir:13a9ec974f6dc2a0279ca6d9c0525d3c4c51936c
                   ;origin=https://doi.org/10.5281/zenodo.19351493;vi
                   sit=swh:1:snp:85311dfbf8c480c19858961737034730ca52
                   3b96;anchor=swh:1:rel:3e5870534c14d78602f6dab07dc3
                   ca291db7929d;path=TKmist-FcsIT-54184b8
                  },
    note      = {Version 1.0.2, Zenodo, Software, DOI:10.5281/zenodo.19367803. (The development repository is hosted on
GitHub: \url{https://tkmist.github.io/FcsIT/}.)},
}
